\documentclass[10pt,aps,twocolumn,amsmath,amssymb,nofootinbib,superscriptaddress,showkeys,showpacs]{revtex4-1}
\usepackage{epsfig} 
\usepackage{amsmath} 
\usepackage{graphicx}
\usepackage{color}
\usepackage{float}
\usepackage{subfigure}

\newcommand{\ba}{\begin{array}}
\newcommand{\ea}{\end{array}}
\newcommand{\bea}{\begin{eqnarray}}
\newcommand{\eea}{\end{eqnarray}}
\newcommand{\be}{\begin{equation}}
\newcommand{\ee}{\end{equation}}

\renewcommand{\l}{\left}
\renewcommand{\r}{\right}

\begin{document} 

\title{Quantum Fisher and Skew information for Unruh accelerated Dirac qubit}

\author{Subhashish Banerjee}
\email{subhashish@iitj.ac.in}
\affiliation{Indian Institute of Technology Jodhpur, Jodhpur 342011, India}

\author{Ashutosh Kumar Alok}
\email{akalok@iitj.ac.in}
\affiliation{Indian Institute of Technology Jodhpur, Jodhpur 342011, India}

\author{S. Omkar}
\email{omkar.shrm@gmail.com }
\affiliation{Indian Institute of Science Education and Research, Thiruvananthapuram, India }

\date{\today} 

\begin{abstract}
We develop a Bloch vector representation of Unruh channel for a Dirac field mode. 
This is used to provide a unified, analytical treatment of quantum Fisher and Skew information 
for a  qubit subjected to the Unruh channel, both in its pure form as 
well as in the presence of experimentally relevant external noise channels.
The time evolution of Fisher and Skew information is studied along with 
the impact of external environment parameters such as temperature and squeezing. 
The external noises are modelled by both purely dephasing phase damping as well as the squeezed generalized
amplitude damping channels. An interesting interplay between the external reservoir temperature and squeezing on the Fisher and Skew information is observed, 
in particular, for the action of the squeezed generalized amplitude damping channel. It is seen that for some regimes, 
squeezing can enhance the quantum information against the deteriorating influence of the ambient environment. 
Similar features are also observed for the analogous study of Skew information, highlighting the similar origin of the Fisher and Skew information.

\end{abstract}

\maketitle 

\section{Introduction}

The  Unruh effect \cite{Dav-Unr,crispino} predicts
that the  Minkowski vacuum  as seen by  an observer  accelerating uniformly
will  appear as a warm  gas emitting black-body radiation at the \textit{Unruh temperature}. The  Unruh effect produces 
a decoherence-like  effect \cite{Omkar:2014hba}.   It degrades  the quantum  information
shared  between  an  inertial  observer   and  an  accelerated
observer, as seen in the  latter's frame, in the case of bosonic
or  Dirac field  modes  \cite{AM03, AFM+2006,  Tian2012}. The studies on Unruh effect form a part of the endeavour 
to understand relativistic aspects of quantum information \cite{czachor1997,peres1,caban1,
Bruschi:2012uf,Friis:2012cx,Lee:2014kaa,Banerjee:2014vga,Alok:2014gya,Banerjee:2015mha}, see for example the review
\cite{Peres:2002wx}.

In this work we take up the problem of studying Fisher and its variant Skew information \cite{fisher,cramer,frieden} for 
the Unruh effect on a Dirac field mode  in the context of open quantum systems \cite{bruer,alicki,weiss}. 
The Fisher information plays a key role
in the estimation of unknown state parameters and provides a lower bound on the error of
estimation \cite{helstrom}. Estimation of initial state parameters has been of interest for quiet some time \cite{petz}
and in recent years this approach has been turned towards state estimation in the context of 
open quantum systems \cite{nori,blandeau}. Another variant of the Fisher information is the Skew information,
which is the related to the infinitesimal form of the quantum Hellinger distance \cite{wigner}.
In recent times Skew information  has been shown to satisfy some nice
properties relevant to the coherence in the system \cite{luo,girolami,vedral}.
Both the quantum Fisher and Skew information are two different aspects of the classical Fisher information
in the quantum regime \cite{luo2}, with the Skew and Fisher information being related to the Hellinger 
and Bures distance, respectively \cite{bures,uhlman}. 
These notions have also been used in recent times to provide a diagnostic for the
general evolution of the quantum system, that is, whether the dynamics is Markovian or non-Markovinan \cite{nonmarkov}.

Here, we develop a  Bloch vector representation characterizing the Unruh channel acting on a qubit, 
to provide analytical expressions for quantum Fisher and Skew information, 
both with and without external noises. For the external noises, we take the experimentally 
relevant \cite{haroche,turchette} purely dephasing QND (Quantum non-demolition) \cite{sbrg} as well as 
the squeezed generalized amplitude damping (SGAD) noise \cite{sbsrik,sbomkar}. The QND channel is a 
purely quantum effect incorporating decoherence without dissipation, while the SGAD channel is a very 
general noisy channel in that it incorporates both the effects of finite temperature and bath squeezing. 
We observe the non-trivial interplay between temperature and bath squeezing on the Fisher and Skew information.
In particular, it is observed that in some regimes squeezing can play a constructive role in enhancing the information
against the deteriorating influence of temperature.

Plan of the work is as follows. In Sec. II we briefly discuss the importance of quantum Fisher information in the context of 
estimation theory and motivate the use of the Bloch vector formalism for the study of  Unruh effect with(out) external noises. 
We then develop the Bloch vector formalism characterizing the Unruh channel. In the next section quantum Fisher information for
the Unruh channel without any external noise is studied. In Sec. V we extend the above by incorporating the effect of external 
noises, both the purely dephasing phase damping as well as SGAD channels. Since the Skew information is another variant of the
quantum Fisher information, we probe Skew information both for the pure Unruh channel as well as for the cases where the channel 
is affected by external noises, QND as well as SGAD. Finally we make our conclusions.

\section{Quantum Fisher Information in the Bloch vector formalism}
 
 With the advent of experimental progress, estimation theory has become a powerful tool for activities 
 such as state reconstruction, tomography and metrology \cite{paris}. Quantum Fisher information plays a prominent role in these
 activities where a question of central importance is the determination of an unknown parameter characterizing the system
 and to reduce the error in these estimations. Their roots are related to the famous Cramer-Rao bounds \cite{cramer,rao} which is related to
 the fundamental founds on the efficiency of the estimation problem. Quantum Fisher information is the quantum counterpart of these
 bounds. 
 
 It is well known that Unruh process is an inherently noisy one. Thus it is of interest to have an understanding of this process
 from the prospective of the estimation problem and hence the motivation for a study of Unruh effect using Fisher information. Efforts have been
 made along these directions,  for example,  in \cite{nori} where a systematic study was made of the problem of Fisher information in the
 presence of a number of well know noisy channels such as the phase damping, which is essentially a QND channel and generalized amplitude damping (GAD)
 channel, which is a subset of the SGAD channel.  A similar study was also made in \cite{blandeau} where the problem of estimation of probe states
 with the feature of best resistance to noise was studied. In both of these works, the geometric visualization offered by the Bloch vector formalism was made use of 
 in estimating the Fisher information. 
 
 In \cite{yao} these studies were applied to the problem of pure Unruh effect both for the scalar as well as the Dirac field mode. Here, using the Kraus operators characterizing the pure Unruh channel, we develop a Bloch vector treatment of the quantum Fisher information of the Unruh effect of a Dirac field mode. Further, we study the effect of external noises, both QND as well as SGAD, on the Fisher estimation. This enables us to study the interplay between the external temperature and reservoir squeezing, a quantum correlation, on the Fisher information. 
Our constructions are analytic in nature. Quantum Fisher information in terms of Bloch vector $\vec{\zeta}(\alpha)$ is given by \cite{nori,blandeau}
\be 
F_q(\alpha) = \frac{\l[\vec{\zeta}(\alpha) \cdot \partial_{\alpha}\vec{\zeta}(\alpha) \r]^2}{1-|\vec{\zeta}(\alpha)|^2} + \l[\partial_{\alpha}\vec{\zeta}(\alpha)\r]^2\,,
\label{fisher}
\ee
where $q$ denotes quantum and $\alpha$ is the parameter to be estimated, for example, the polar and azimuthal angles $\theta$ and $\phi$, respectively, of a qubit.
From here on we will abbreviate $F_q(\alpha)$ by $F_{\alpha}$.
We will make use of this in our work.

\section{Bloch vector formalism for Unruh channel}

Here we provide a sketch of the tools required to obtain a Bloch vector representation of the Unruh channel. 
The basic ingredient that goes into this endeavour is the Choi theorem \cite{choi,debbie} which is applied here
by considering  the  maximally entangled two Dirac field modes state $\frac{1}{\sqrt{2}}(|00\rangle + |11\rangle)$ in which  the second
mode  is Unruh  accelerated. This results in   \cite{Omkar:2014hba}  
\begin{equation} 
\rho_U=\frac{1}{2}\left(
 \begin{array}{cclr} \cos^2r&0&0&\cos r\\ 0&\sin^2r&0&0\\ 0&0&0&0\\
\cos r&0&0&1
 \end{array}\right),
\label{eq:uchoi}
\end{equation} 
 where $r$ is the Unruh parameter given by $\cos
r=\frac{1}{\sqrt{e^{-\frac{2\pi\omega   c}{a_u}}+1}}$. Here $a_u$ is the uniform Unruh acceleration
and $\omega$ is the Dirac particle frequency.
As $a_u$ ranges from $\infty$  to 0, $\cos  r \in[\frac{1}{\sqrt{2}},1]$.
The spectral decomposition of the above state gives 
\begin{equation}
\rho_U=\sum_{j=0}^3|\xi_j\rangle\langle\xi_j|,
\label{spectral} 
\end{equation}  
where $|\xi_j\rangle$ are the eigenvectors  normalized to the value of
the  eigenvalue.   Choi's  theorem  \cite{choi,debbie}, by making use of channel-state duality, 
then provides a root to obtaining the
Kraus operators relevant to the channel generating the state in Eq.~\ref{eq:uchoi}. Essentially,
each $|\xi_j\rangle$ yields a Kraus operator  obtained by folding the $d^2$
 elements of the  eigenvector into a $d\times  d$ 
matrix,   by taking each sequential  $d$-element segment of
$|\xi_j\rangle$, writing  it as a  column, and then  juxtaposing these
columns to form the matrix \cite{debbie}. Here $d=2$.

Spectral decomposition of $\rho_U$, Eq.~\ref{spectral}, yields the following  eigenvectors,  corresponding to two non-vanishing eigenvalues
\bea
|\xi_0\rangle&=&(\cos r,0,0,1),\nonumber\\
|\xi_1\rangle&=&(0,\sin r,0,0).
\eea
A straightforward application of Choi's theorem now yields  the  following Kraus operators  for  the Unruh channel $\mathcal{E}_U$ as  
\be
{\mathcal K}^U_1=\left(
\begin{array}{cclr}
\cos r&0\\
0&1
\end{array}\right);~~
{\mathcal K}^U_2=\left(
\begin{array}{cclr}
0&0\\
\sin r&0
\end{array}\right),
\label{eq:relcha}
\end{equation}
whereby
\begin{equation}
\mathcal{E}_U(\rho) = \sum_{j=1,2} {\mathcal K}^U_j \rho \left({\mathcal K}^U_j\right)^\dag,
\label{eq:unrucha}
\end{equation}
with the completeness condition
\begin{equation}
\sum_{j=1,2} \left({\mathcal K}^U_j\right)^\dag  {\mathcal K}^U_j = \mathbb{I}.
\end{equation}

From the above Kraus representation, it would appear that the Unruh channel is formally similar to
an AD  channel \cite{NC00}, which models the effect of a zero temperature bath \cite{NC00,sbsrik,sbrg}.  This
  is  surprising  as   the  Unruh  effect  corresponds   to  a  finite
  temperature  and would  naively  be expected  to  correspond to   finite
  temperature channels such as the GAD   or  SGAD   channels. 
  
Any two level system can be represented in the Bloch vector formalism as
\be
\rho = \frac{1}{2} \l(     \mathbb{I}+ \vec{\zeta} \cdot \sigma \r)\,,
\label{bloch}
\ee
where $\sigma$ are the standard Pauli matrices.
For the initial state $\rho=
|0\rangle\langle0|\cos^2\frac{\theta}{2}
+ |0\rangle\langle1|e^{i\phi}\cos\frac{\theta}{2}\sin\frac{\theta}{2} +
|1\rangle\langle0|e^{-i\phi}\cos\frac{\theta}{2}\sin\frac{\theta}{2}
+ |1\rangle\langle1|\sin^2\frac{\theta}{2},$ the Bloch vector can be seen to be
$\zeta_0 = \l(\cos \phi \sin\theta,\, -\sin \phi \sin \theta,\, \cos \theta\r)$. 
Evolving this state under the Unruh channel, characterized by the above
Kraus operators leads to  a state, which could be called the Unruh-Dirac (UD) qubit state,  whose Bloch vector is 
\begin{equation}
\vec{\zeta}= \left(\begin{array}{clr}
\cos r \cos \phi \sin \theta \\
\vspace{.2cm}
- \cos r \sin \phi \sin \theta \\
\vspace{.2cm}
\cos^2r \cos\theta - \sin^2r
\end{array}\right)=A \vec{\zeta_0}+C.
\label{bloch-unruh}
\end{equation}

From this $A$ and $C$ can be found to be
\be 
A= \left(\begin{array}{clclclr}
\cos r & 0 & 0\\
\vspace{.2cm}
0 & \cos r & 0 \\
0 & 0 & \cos^2 r
\end{array}\right), \quad 
C= \left(\begin{array}{clr}
0 \\
\vspace{.2cm}
0\\
\vspace{.2cm}
-\sin^2 r 
\end{array}\right).
\label{AC}
\ee
This, we believe, is a new result with the  $A$ and $C$ matrices completely 
characterizing the Unruh channel and will be used in the  investigations below. 

\section{Quantum Fisher information for Unruh channel without external noise}

That the Unruh channel is an inherently noisy channel is made explicit by its Kraus representation Eqs.~(\ref{eq:relcha}) and (\ref{eq:unrucha}). 
Here we will estimate the UD qubit state using quantum Fisher information. For this purpose we will make use of $A$ and $C$ from 
Eq.~(\ref{AC}) and $\vec{\zeta}$ from Eq.~(\ref{bloch-unruh})
as inputs in Eq.~(\ref{fisher}).

When no external noise is acting, i.e., for the case of the pure Unruh channel, it can be seen that
\bea
F_\theta&=&\cos^2r,\nonumber\\
F_\phi&=&\cos^2r\sin^2\theta.
\label{pureunruh}
\eea
The Fisher information with respect to the parameter $\theta$, $F_\theta$, is independent of the state parameter $\theta$ while  
the Fisher information with respect to the parameter $\phi$, $F_\phi$ is state dependent and depends upon $\theta$. It should be noted that
both these expressions of Fisher information have no $\phi$ dependence. Also it can be observed from the above expressions that the Fisher information cannot be 
increased by increasing the Unruh acceleration. This is consistent with the fact that the Unruh acceleration produces a thermal like effect and
quantum estimation would be expected not to increase with increase in temperature. 

\begin{figure*}[ht]
\subfigure[]{
    \includegraphics[width=0.48\textwidth]{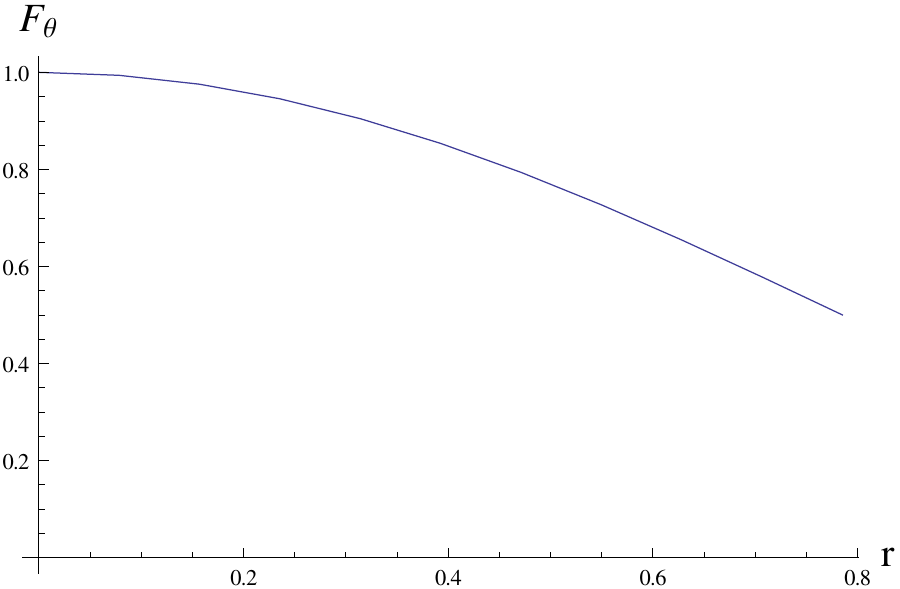}}
\subfigure[]{
 \includegraphics[width=0.48\textwidth]{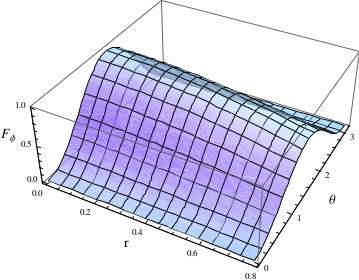}}
\caption{(a) Plot of $F_\theta$ with respect to Unruh parameter for acceleration $r$; 
(b) Plot of $F_\phi$ with respect to $r$ and $\theta$.}
\label{fig:without-noise}
\end{figure*}

This is also evident from Fig.~\ref{fig:without-noise}, where $F_{\theta}$ is plotted as a function of the Unruh parameter $r$ whereas $F_{\phi}$, is plotted with respect to $r$ and $\theta$. 
As $r$ goes from $\pi/4$ to 0, i.e., $\cos r$ goes from $1/\sqrt{2}$ to 1, which implies the Unruh acceleration $a$ decreasing from infinity to zero, $F_{\theta}$ increases to 1. Since  the Unruh acceleration is directly proportional to temperature,  as acceleration decreases, temperature also decreases and quantum Fisher information increases. This is also seen for $F_\phi$, albeit only for $\theta=\pi$.

Further, it can be seen from above that $F_\theta$ depends upon the Unruh parameter $r$. It should be noted that 
this result is obtained for an Unruh channel, by accelerating one partner of the maximally entangled state, as indicated in the previous section. It is
interesting to observe here that for an analogous study of Unruh effect on a different state, not necessarily maximally entangled, $F_\theta$
was shown to be independent of $r$ \cite{yao}. This suggests that Fisher information which 
is an important tool in  state estimation could also be used as a witness for quantum correlations.  

\section{Quantum Fisher information for Unruh channel with external noise}

Now we will analyze the effect of external noise  on the Unruh channel using 
the Bloch vector formalism of Quantum Fisher information. For this, we consider two 
general external noisy channels : a) phase damping channel, which is of the QND 
kind and involves pure dephasing and  b) the SGAD channel, which includes the effects of decoherence
along with dissipation and accounts for finite bath temperature as well as squeezing.  
 We adopt the following procedure. Starting 
from the UD qubit state, Eq.~(\ref{bloch-unruh}),  application 
of the external noise channel results in
\be 
\rho_{\rm in} \xrightarrow{\mathcal{E}{\rm(phase/SGAD)}} \rho_{\rm new}\,.
\ee
 From $\rho_{\rm new}$, we get the new Bloch vector $\vec{\zeta}_{\rm new}$ which is related to the original state Bloch vector as 
\bea 
\vec{\zeta}_{\rm new} &=& A' \vec{\zeta} +C' = A'(A\vec{\zeta_0}+C)+C'\nonumber\\
&=&AA'\vec{\zeta_0} + (A'C+C')\equiv A_{\rm new}\vec{\zeta_0}+C_{\rm new}.
\label{zetanew}
\eea 
Here $\vec{\zeta}$ and $\vec{\zeta_0}$ are as in Eq.~(\ref{bloch-unruh}).
From the above equation, it can be seen that the effect of the external noise channel on the Unruh channel is encoded in
 $A_{\rm new}=A'A$ and $C_{\rm new} = (A'C + C')$.  Thus we need to find out
$A'$ and $C'$ for the desired channels using the Kraus operator formalism.

\subsection{ Phase damping channel}
In the context of open quantum systems, one is interested in the dynamics of the system of interest, for example, the UD qubit in this case, by taking into account the effect of the ambient environment on its evolution. Let the total Hamiltonian $H$ be $H=H_S + H_R + H_{SR}$, where $H_S$, $H_R$ are the system and reservoir Hamiltonians, respectively and $H_{SR}$ is the interaction between the two. If $[H_S, H_{SR}]=0$, then it implies
decoherence without dissipation, that is, pure dephasing. This is a purely quantum mechanical effect and such an interaction is called a QND interaction. The phase damping channel is a well known noisy channel incorporating QND interaction.

The Kraus operators corresponding to the phase damping channel,  modelling the QND interaction of a qubit, with the two levels having a separation of $\hbar \omega_0$, interacting with a squeezed thermal bath are \cite{sbrg} 
\bea
K_1&=&\sqrt{\frac{1+e^{-(\hbar\omega_0)^2\gamma(t)}}{2}}\left(\begin{array}{cc}
e^{-i\hbar\omega_0 t}&0\\
0&1
\end{array}\right);\nonumber\\
K_2&=&\sqrt{\frac{1-e^{-(\hbar \omega_0)^2\gamma(t)}}{2}}\left(\begin{array}{cc}
-e^{-i\hbar \omega_0 t}&0\\
0&1
\end{array}\right).
\eea
Assuming an Ohmic bath  spectral density with an upper cut-off frequency $\omega_c$, it can be shown that
\begin{widetext}
\bea
\gamma(t)&=&  \left(\frac{\gamma_0 k_B T}{\pi \hbar \omega_c}\right)
 \cosh(2 s) \left(2 \omega_c t\tan^{-1}(\omega_c t) + 
\ln\left[\frac{1}{1 + \omega_c^2 t^2} \right]\right) - \left(\frac{\gamma_0 k_B T}{2\pi  \hbar \omega_c}\right) \sinh(2 s) \Bigg(\frac{}{}4 \omega_c(t - a) \tan^{-1}[2 \omega_c (t - a)]\nonumber\\
&&
-4 \omega_c (t-2 a) \tan^{-1}[\omega_c (t-2a)] 
+  4 a \omega_c \tan^{-1}(2 a \omega_c) + \ln\left[\frac{\left(1 + \omega_c^2 (t - 2 a)^2\right)^2}{1 + 4\omega_c^2(t-a)^2}\right] + \ln\left[\frac{1}{1 + 4 a^2\omega_c^2}\right]\Bigg).
\eea
\end{widetext}
Here $T$ is the reservoir temperature, while $a$ and $s$ are bath squeezing parameters. 
For the Unruh channel in the presence of phase damping noise the modified Bloch vector for the UD qubit, Eq.~(\ref{zetanew}), is
\begin{equation}
\zeta_{\textrm{new}}=
\begin{pmatrix}
\cos r\sin\theta\cos(\phi+\omega_0 t)e^{-(\hbar\omega_0)^2\gamma(t)/4}\\
-\cos r\sin\theta\sin(\phi+\omega_0 t)e^{-(\hbar\omega_0)^2\gamma(t)/4}\\
\cos^2r \cos\theta - \sin^2r
\end{pmatrix}.
\end{equation}

\begin{figure*}[ht]
\subfigure[]{
\includegraphics[width=0.48\textwidth]{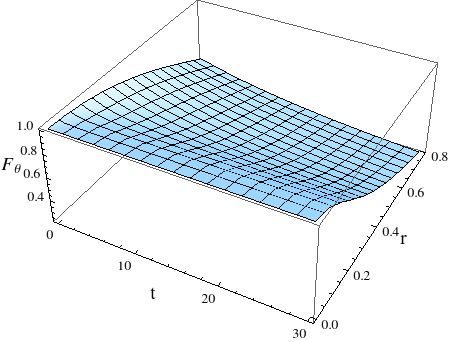}}
\subfigure[]{
\includegraphics[width=0.48\textwidth]{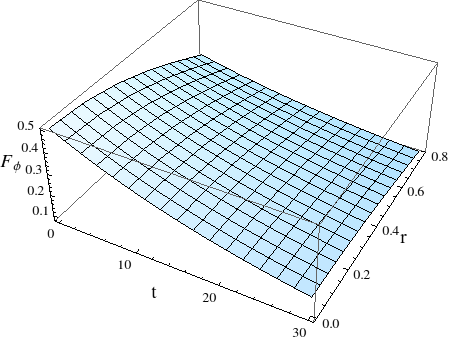}}
\caption{(a) Variation of $F_\theta$ (Fisher information with respect to the  parameter $\theta$) and (b) $F_\phi$ (Fisher information information with respect to parameter $\phi$) for QND interaction with bath for a time  (t) and and Unruh parameter (r).
The parameter settings are  $\theta=\pi/4$, $\phi=\pi/4$, $a=0$, $T=0.5$, $s=0.5$, $\omega_0=1$, $\omega_c=100$, $\gamma_0=0.1$.}
\label{qndfisher}
\end{figure*}

\begin{figure*}[ht]
\centering
\subfigure[]{
\includegraphics[width=0.48\textwidth]{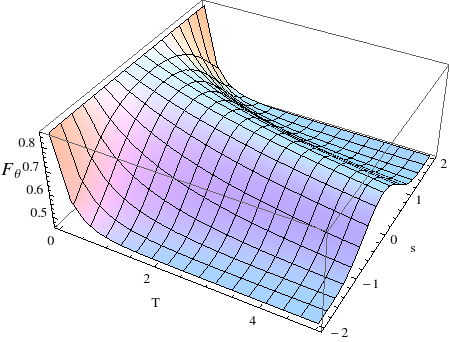}}
\subfigure[]{
\includegraphics[width=0.48\textwidth]{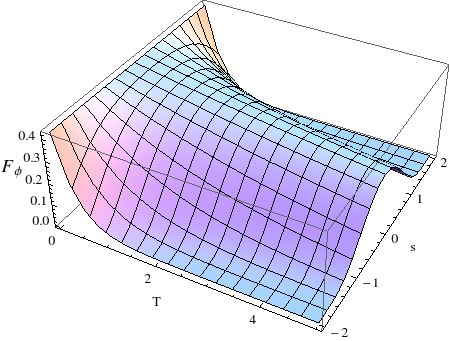}}
\caption{(a) Variation of $F_\theta$ (Fisher information with respect to parameter $\theta$) for QND interaction with bath temperature (T) and squeezing (s); 
(b) Variation of $F_\phi$ (Fisher information with respect to parameter $\phi$) for QND interaction with bath temperature (T) and squeezing (s).
The parameter settings are $r=\pi/8$, $\theta=\pi/4$, $\phi=\pi/4$, $a=0$, $\omega_c=100$, $\gamma_0=0.1$, $t=2$.}
\label{fisherpd}
\end{figure*}

The analytical expressions for the quantum Fisher information with respect to parameters $\theta$ and $\phi$ are
\begin{widetext}
\bea
F_\theta &=& \frac{\cos^2 r \left(2(1+\cos2r) + \cos(2 r-\theta) + 4 e^{\frac{\gamma(t)  (\hbar\omega_0) ^2}{2}}(\cos\theta-1)-6\cos\theta + \cos(2r+\theta) \right)}
{4(1-\cos\theta) + 2e^{\frac{\gamma(t)  (\hbar\omega_0) ^2}{2}}(\cos2r  -3 + 2\cos^2r\cos\theta)},\nonumber\\
F_\phi &=& e^{-\frac{\gamma(t)  (\hbar\omega_0) ^2}{2}} \cos^2r \sin^2\theta,
\eea
\end{widetext}
respectively. The above expressions of the Fisher information reduce, for $\gamma(t)=0$, to their pure Unruh counterparts in Eq.~(\ref{pureunruh}). 
Also, both $F_{\theta}$ and $F_{\phi}$ are independent of the azimuthal angle $\phi$, as in the pure Unruh case. 

From Fig.~\ref{qndfisher} it can be seen that both $F_\theta$ and $F_\phi$ 
decrease with time with increase in Unruh acceleration parametrized by $r$.
However, for $r<0.2$, $F_\theta$ is stable with the evolution of time.
In Fig.~\ref{fisherpd} profiles of $F_{\theta}$ and $F_{\phi}$  with respect to $T$ 
and $s$ are depicted. It is evident that the Fisher information decreases with 
increasing $T$. Also, squeezing is seen to have a depleting effect on the  Fisher 
information.

\subsection{ SGAD channel}
Usually, $[H_S, H_{SR}]\neq0$, implying deoherence along with dissipation. 
Linbladian evolution \cite{lindblad,sudharshan,alicki} is a general class of evolutions which incorporates the effects of decoherence and dissipation.
The SGAD channel is a very general Linbladian noisy channel incorporating the effects of bath squeezing, dissipation
and decoherence. The Kraus operators for this  channel are \cite{sbsrik,sbomkar}
\begin{eqnarray}
K_{1} &\equiv& \sqrt{p_1}\left[\begin{array}{ll} 
\sqrt{1-\alpha} & 0 \\ 0 & 1
\end{array}\right], \nonumber\\
K_{2} &\equiv& \sqrt{p_1}\left[\begin{array}{ll} 0 & 0 \\ \sqrt{\alpha} & 0
\end{array}\right],  \nonumber\\
K_{3} &\equiv& \sqrt{p_2}\left[\begin{array}{ll} \sqrt{1-\mu} & 0 \\ 0 & 
\sqrt{1-\nu}
\end{array}\right], \nonumber\\
K_{4} &\equiv& \sqrt{p_2}\left[\begin{array}{ll} 0 & \sqrt{\nu} \\ \sqrt{\mu}e^{-i\phi_s} & 0
\end{array}\right],
\label{eq:srikraus}
\end{eqnarray}
where $p_1+p_2=1$ \cite{sbsrik}, and
\begin{widetext}
\begin{eqnarray}
p_2 &=& \frac{1}{(A+B-C-1)^2-4D}
 \times \left[A^{ 2} B + C^{2} + A(B^{2} - C - B(1+C)-D) - (1+B)D - C(B+D-1)
\nonumber \right. \\
&& \pm \left.  2\sqrt{D(B-A B+(A-1)C+D)(A-A B+(B-1)C+D)}\right],
\label{eq:p2}
\end{eqnarray}
\end{widetext}
with
\begin{eqnarray}
A &=& \frac{2N+1}{2N} \frac{\sinh^2(\gamma_0 at/2)}
{\sinh(\gamma_0(2N+1)t/2)}
\exp\left(-\gamma_0(2N+1)t/2\right),\nonumber \\
B &=& \frac{N}{2N+1}(1-\exp(-\gamma_0(2N+1)t)), \nonumber \\
C &= & A + B + \exp(-\gamma_0 (2N+1)t),\nonumber \\
D &=& \cosh^2(\gamma_0 at/2)\exp(-\gamma_0(2N+1)t).
\label{eq:auxip2}
\end{eqnarray}
Also,
\begin{eqnarray}
\nu &=& \frac{N}{(p_2)(2N+1)}(1-e^{-\gamma_0(2N+1)t}),\nonumber \\
\mu &=& \frac{2N+1}{2(p_2) N}\frac{\sinh^2(\gamma_0at/2)}{\sinh(\gamma_0(2N+1)t/2)}
\exp\left(-\frac{\gamma_0}{2}(2N+1)t\right),\nonumber \\
\alpha & =& \frac{1}{p_1}\left(1 - p_2[\mu(t)+\nu(t)]
- e^{-\gamma_0(2N+1)t}\right).
\label{eq:nu}
\end{eqnarray} 
Further,
$ N = N_{\rm th}[\cosh^2(s) + \sinh^2(s)]  + \sinh^2(s),~~a=\sinh(2s)( 2N_{\rm th}+1)$ where 
$N_{\rm th}= 1/(e^{\hbar  \omega_0/k_B T} -  1)$ is the Planck  distribution giving
the  number of  thermal photons  at  the frequency  $\omega_0$ while $s$  and
$\phi_s$  are  bath  squeezing  parameters.

\begin{figure*}[ht]
\subfigure[]{
\includegraphics[width=0.48\textwidth]{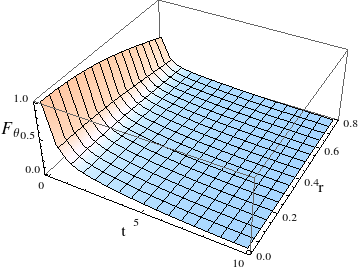}}
\subfigure[]{
\includegraphics[width=0.48\textwidth]{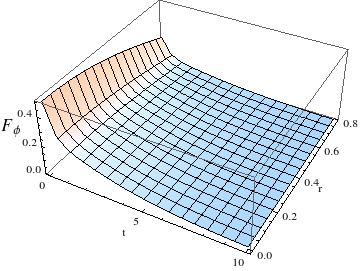}}
\caption{(a) Variation of $F_\theta$ (Fisher information with respect to the  parameter $\theta$) and (b) $F_\phi$ 
(Fisher information information with respect to parameter $\phi$) for SGAD interaction with bath interaction time (t) 
and and Unruh parameter (r). The parameter settings are $T=0.5$, $s=0.5$, $\theta=\pi/4$, $\phi=\pi/4$, $\phi_s=0$, $\omega_0=0.1$,             $\gamma_0=0.1$.}
\label{sgadskew}
\end{figure*}

Under the action of the SGAD channel, the effective Bloch vector Eq.~(\ref{zetanew}) becomes
\begin{widetext}
\begin{equation}
\zeta_{\textrm{new}}=
\begin{pmatrix}
\cos r\sin\theta\left(\left(p_1\sqrt{1-\alpha }+p_2\sqrt{(1-\mu )(1-\nu )}\right)\cos\phi+p_2\sqrt{\mu\nu}\cos(\phi-\phi_s)\right)\\
-\cos r\sin\theta\left(\left(p_1\sqrt{1-\alpha }+p_2\sqrt{(1-\mu )(1-\nu )}\right)\sin\phi-p_2\sqrt{\mu  \nu }\sin(\phi-\phi_s)\right)\\

\left(1-2p_1\alpha -2p_2\mu \right)\cos^2r\cos^2\frac{\theta}{2}-\left(1-2p_2\nu \right)\left(\sin^2r\cos^2\frac{\theta}{2}+\sin^2\frac{\theta}{2}\right)\\
\end{pmatrix}.
\end{equation}
\end{widetext}
As a result, the Fisher information with respect to the Unruh qubit parameters 
$\theta$ and $\phi$, i.e., $F_\theta$ and $F_\phi$, respectively, can be shown to be

\begin{widetext}
\bea
F_\theta&=&\frac{\cos^2\theta(\mathcal{A}_+^2+\mathcal{B}_+^2)+\sin^2\theta~\mathcal{C}^2
+\left(\mathcal{C}\mathcal{D}+ (\mathcal{A}_+ + \mathcal{B}_+)\cos\theta\right)^2\sin^2\theta}
{1- (\mathcal{F} - \mathcal{C}\cos^2\frac{\theta }{2})^2 -(\mathcal{A}_+^2+\mathcal{B}_+^2)\sin^2\theta},\nonumber\\
F_\phi&=&\frac{\sin^2\theta(\mathcal{A}_-^2+\mathcal{B}_-^2)+(\mathcal{A}_+\mathcal{B}_-+\mathcal{A}_-\mathcal{B}_+)^2\sin^4\theta}
{1- (\mathcal{F} - \mathcal{C}\cos^2\frac{\theta }{2})^2 -(\mathcal{A}_+^2+\mathcal{B}_+^2)\sin^2\theta}.
\label{fis-sgad}
\eea
\end{widetext}
Here

\bea
\mathcal{A}_\pm&=&\left[\left(p_1 \sqrt{1-\alpha }+p_2 \sqrt{(\mu-1 ) (\nu-1 )}~\right) \sin\phi\right.  \nonumber\\
&& \left.\mp p_2 \sqrt{\mu  \nu } \sin[\phi -\phi_s]\frac{}{}\right]\cos r,\nonumber\\
\mathcal{B}_\pm&=&\left[\left(p_1 \sqrt{1-\alpha }+p_2 \sqrt{(\mu-1 ) (\nu-1 )}~\right) \cos\phi\right.  \nonumber\\
&& \left.\pm p_2 \sqrt{\mu  \nu } \cos[\phi -\phi_s]\frac{}{}\right]\cos r,\nonumber\\
\mathcal{C}&=&\left[ p_1 (\alpha-1 )+p_2 (2 \nu-1 )\right]\cos^2r,\nonumber\\
\mathcal{D}&=& (1-2 p_1 \alpha -2 p_2 \mu ) \cos^2r \cos^2\frac{\theta }{2} \nonumber\\
&&- (1-2 p_2 \nu ) \left(\cos^2\frac{\theta }{2}\sin^2r+\sin^2\frac{\theta }{2}\right),\nonumber\\
\mathcal{F}&=&-(p_1\alpha+p_2\mu)\cos^2r\cos^2\frac{\theta }{2}\nonumber\\
&&     +\frac{1}{2}(1-2p_2\nu)(-1+\cos\theta-2\sin^2r\frac{\theta }{2}). 
\label{eq:reduction}
\eea

In the absence of external noise $F_\theta$ and $F_\phi$, Eq.~(\ref{fis-sgad}), reduce to Eq.~(\ref{pureunruh}), corresponding to the pure Unruh channel. 

\begin{figure*}[ht]
\subfigure[]{
\includegraphics[width=0.48\textwidth]{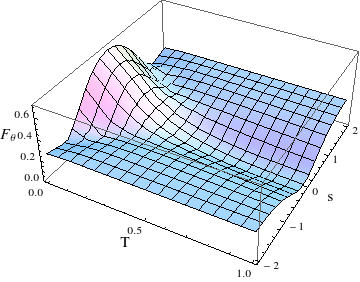}}
\subfigure[]{
\includegraphics[width=0.48\textwidth]{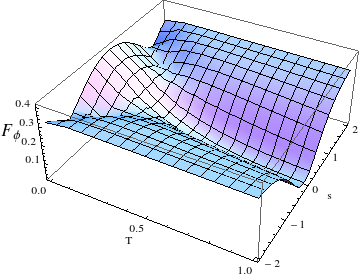}}
\caption{(a) Variation of $F_\theta$ (Fisher information with respect to parameter $\theta$) for SGAD interaction 
with bath temperature (T) and squeezing (s); 
(b) Variation of $F_\phi$ (Fisher information with respect to parameter $\phi$) for SGAD interaction
 with bath temperature (T) and squeezing (s).
The parameter settings are $r=\pi/8$, $\theta=\pi/4$, $\phi=\pi/4$, $\phi_s=0$, $\omega_0=0.1$, $\gamma_0=0.1$, $t=2$.}
\label{fishersgad}
\end{figure*}

From Fig.~\ref{sgadskew}, it is evident that both $F_\theta$ and $F_\phi$ decrease with time for all values of $r$. 
This is in contrast with the corresponding behaviour of $F_\theta$ in the presence of phase damping channel.
The fall in $F_\phi$ here is more dramatic as compared to its phase damping counterpart. 

The variation of $F_\theta$ and $F_\phi$ with respect to temperature and squeezing are shown in Fig.~\ref{fishersgad}. Fisher information
shows an interesting behaviour with respect to squeezing. For a given $T>0.5$, as $s$  becomes nonzero, both $F_\theta$ and $F_\phi$
show a general trend of increasing and stabilizing after $|s|=1$.  To summarize, here squeezing turns out to be a useful quantum resource in that it quantifies the resilience of the quantum system to the effects of the external noisy channel. 

Another feature that is observed is that as the temperature due to the external noise channel increases, the pattern of $F_\phi$
with respect to the Unruh parameter $r$ and state parameter $\theta$, as seen in Fig.~\ref{fig:without-noise}(b), remains unchanged although the magnitude of $F_\phi$  decreases, with the
depletion being more dramatic for the case of the SGAD channel as compared to the QND channel.
 
\section{Skew Information}

Another variant of Fisher information which accounts for  the amount of information in the quantum state with respect to its
non commutation with a conserved quantity is the Skew information \cite{wigner,luo2,nori,pires}. This can be shown to have a metrical structure given by
the quantum Hellinger distance \cite{luo2} which in turn is related to the quantum affinity and is intrinsically connected to the  quantum Chernoff distance \cite{wei14}. 
In this sense Skew and Fisher information are variants of the same fundamental quantity with Fisher deriving its metrical origin from the Bures distance \cite{uhlman}.
Recently there has been a lot of activity concerning the connection between the Skew information and quantum coherence \cite{girolami,vedral}.
We thus find it instructive to compute the Skew information  for the present problem of Unruh channel with and without the influence of external noisy channels.

The Skew information in terms of Bloch vector $\vec{\zeta}(\alpha)$ is given by
\bea 
S_q(\alpha) &=& \frac{2|\partial_{\alpha}\vec{\zeta}(\alpha)|^2}{1+\sqrt{1-|\vec{\zeta}(\alpha)|^2}} + \left[\vec{\zeta}(\alpha)\cdot\partial_{\alpha}\vec{\zeta}(\alpha)\right]^2 \nonumber\\
&& \times \left( \frac{1}{1-|\vec{\zeta}(\alpha)|^2}-\frac{1}{1+\sqrt{1-|\vec{\zeta}(\alpha)|^2}} \right),
 \label{skew}
 \eea
 where $q$ denotes quantum and $\alpha$ is the parameter to be estimated, for example, the polar and azimuthal angles $\theta$ and $\phi$, respectively, of the UD qubit.
 From here on we will abbreviate $S_q(\alpha)$ by $S_{\alpha}$.
 
 \begin{figure*}[ht]
\subfigure[]{
\includegraphics[width=0.48\textwidth]{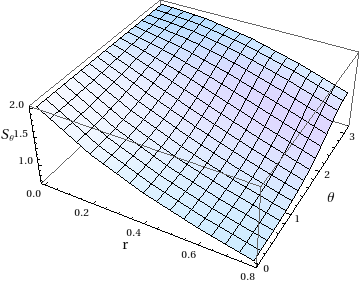}}
\subfigure[]{
\includegraphics[width=0.48\textwidth]{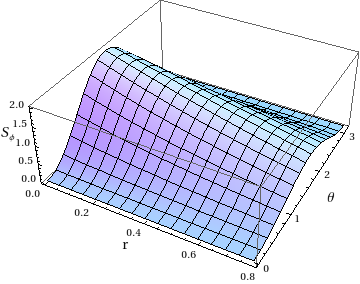}}
\caption{For the pure Unruh channel: (a) Variation of $S_\theta$ (Skew information with respect to the parameter $\theta$); 
(b)  Variation of $S_\phi$ (Skew information with respect to the parameter $\theta$).}
\label{unruhskew}
\end{figure*}

Using the Bloch vector $\vec{\zeta}(\alpha)$, Eq.~(\ref{bloch-unruh}),  the Skew information for the pure Unruh channel with respect to the parameters 
$\theta$ and $\phi$, i.e., $S_\theta$ and $S_\phi$, respectively, can be shown to be
\bea
S_\theta &=& \frac{ \cos^2r \left(7+2 \cos2 \theta + 8 \cos^2\frac{\theta}{2}\sin2r +2\cos2r \sin^2 \theta\right)}
{4\left(1+ \cos^2\frac{\theta}{2}\sin2r\right)^2},\nonumber\\
S_\phi &=& \frac{2 \cos^2r \sin^2\theta}{1+\cos^2\frac{\theta }{2} \sin 2r}.
\label{skew-unr}
\eea
Unlike the analogous case of Fisher information $F_\theta$ for the pure Unruh channel, Eq.~(\ref{pureunruh}), we see that $S_\theta$ depends both on $r$ and $\theta$.

From Fig.~\ref{unruhskew}, it is seen that the Skew information, for the pure Unruh channel,  with respect to the parameter $\theta$, decreases with increase in 
the Unruh parameter $r$, a behaviour which is consistent with that of its Fisher counterpart. However, in contrast to the Fisher information, for a given $r$, there is a general
trend of increase in $S_{\theta}$ as $\theta$ goes from 0 to $2\pi$. This increase is more dramatic for higher values of Unruh acceleration. The behaviour of $S_\phi$ is
similar to that of its Fisher counterpart $F_{\phi}$, Fig.~\ref{fig:without-noise}(b). However for higher values of $r$ ($>0.5$), $S_\phi$ has a steeper fall as compared to its corresponding $F_\phi$, Fig.~\ref{fig:without-noise}.

\subsection{Phase Damping}

\begin{figure*}[ht]
\subfigure[]{
\includegraphics[width=0.48\textwidth]{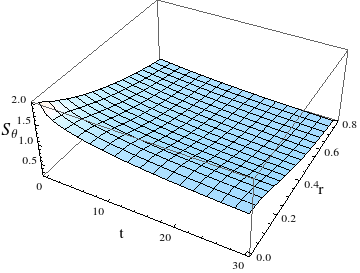}}
\subfigure[]{
\includegraphics[width=0.48\textwidth]{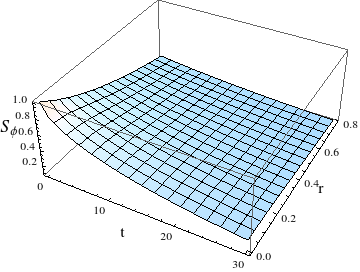}}
\caption{(a) Variation of $S_\theta$ (Skew information with respect to the  parameter $\theta$) and (b) $S_\phi$ (Skew information information with respect to parameter $\phi$) for QND interaction with bath interacting for a time  (t) and and Unruh parameter (r).
The parameter settings are $T=0.5$, $s=0.5$, $\theta=\pi/4$, $\phi=\pi/4$, $a=0$, $\omega_0=1$, $\omega_c=100$, $\gamma_0=0.1$.}
\label{qndskew}
\end{figure*}

The Skew information with respect to parameters $\theta$ and $\phi$, $S_{\theta}$ and $S_{\phi}$, 
due to the influence of the phase damping (QND) noise channel on the UD quibit are given by
\bea
S_\theta &=& \frac{2\cos^2r \left(e^{-\frac{1}{2}\gamma(\hbar\omega_0) ^2}\cos^2\theta+ \cos^2r\sin^2\theta \right)} {1+\sqrt{1-\mathcal{H}}}\nonumber\\
&&+\mathcal{G}^2\sin^2\theta
\left(\frac{1}{1-\mathcal{H}}-\frac{1}{1+\sqrt{1-\mathcal{H}}}\right),\nonumber\\
S_\phi &=& \frac{2 e^{-\frac{\gamma  (\hbar\omega_0) ^2}{2}} \cos^2 r\sin^2\theta}
{1+\sqrt{1-\mathcal{H}}}.
\eea
Here $\mathcal{G}$ and $\mathcal{H}$ are
\bea 
\mathcal{G}&=&e^{-\gamma(\hbar\omega_0) ^2}\cos^4r (\cos\theta+ e^{\frac{1}{2}\gamma(\hbar\omega_0) ^2}(\sin^2r-\cos^2r\cos\theta)),\nonumber\\
\mathcal{H}&=&(\sin^2 r-\cos^2 r \cos\theta)^2-e^{-\frac{\gamma (\hbar\omega_0) ^2}{2}} \cos^2 r \sin^2\theta.
\eea

\begin{figure*}[ht]
\subfigure[]{
\includegraphics[width=0.48\textwidth]{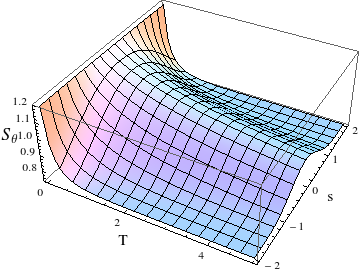}}
\subfigure[]{
\includegraphics[width=0.48\textwidth]{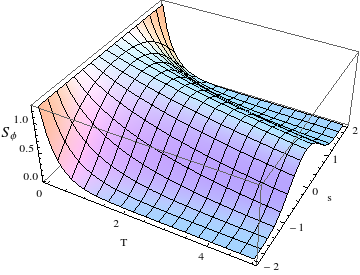}}
\caption{(a) Variation of $S_\theta$ (Skew information with respect to the  parameter $\theta$) and (b) $S_\phi$ (Skew information information with respect to parameter $\phi$) for 
QND interaction with bath at temperature (T) and squeezing (s).
The parameter settings are $r=\pi/8$, $\theta=\pi/4$, $\phi=\pi/4$, $a=0$, $\omega_0=1$, $\omega_c=100$, $\gamma_0=0.1$, $t=2$.}
\label{qndskew-ts}
\end{figure*}
In the absence of external noise, $S_\theta$ and $S_\phi$ reduces to their pure Unruh counterparts in Eq.~(\ref{skew-unr}).

The variation of Skew informations $S_\theta$ 
and $S_\phi$ with respect to time of evolution $t$, Unruh parameter $r$ and temperature (T) and squeezing (s)
are depicted in Fig.~\ref{qndskew} and Fig.~\ref{qndskew-ts}, respectively.   
Once more we see that the behaviour of Skew information is similar to its Fisher counterpart.

\subsection{SGAD}
As a result of action of SGAD channel on UD qubit, the Skew information with respect to parameters $\theta$ and $\phi$ are respectively
given by

\begin{widetext}
\bea
S_\theta&=&\frac{2\left(\cos^2\theta(\mathcal{A}_+^2+\mathcal{B}_+^2)+\mathcal{C}^2\sin^2\theta\right)}
{1+\sqrt{1- (\mathcal{F} - \mathcal{C}\cos^2\frac{\theta }{2})^2 -(\mathcal{A}_+^2+\mathcal{B}_+^2)\sin^2\theta}}
+\left(\mathcal{C}\mathcal{D}\sin\theta +(\mathcal{A}_+^2+\mathcal{B}_+^2)\sin\theta\cos\theta\right)^2\times\nonumber\\
&&\left(\frac{1}{1- (\mathcal{F} - \mathcal{C}\cos^2\frac{\theta }{2})^2 -(\mathcal{A}_+^2+\mathcal{B}_+^2)\sin^2\theta}-\frac{1}{1+\sqrt{1- (\mathcal{F} - \mathcal{C}\cos^2\frac{\theta }{2})^2 -(\mathcal{A}_+^2+\mathcal{B}_+^2)\sin^2\theta}}\right),\nonumber\\
S_\phi&=&\frac{2\left(\mathcal{A}_-^2+\mathcal{B}_-^2\right)\sin^2\theta}
{1+\sqrt{1- (\mathcal{F} - \mathcal{C}\cos^2\frac{\theta }{2})^2 -(\mathcal{A}_+^2+\mathcal{B}_+^2)\sin^2\theta}}
+ \left(\mathcal{A}_+\mathcal{B}_- + \mathcal{A}_-\mathcal{B}_+ \right)^2\sin^4\theta\times\nonumber\\
&&\left(\frac{1}{1- (\mathcal{F} - \mathcal{C}\cos^2\frac{\theta }{2})^2 -(\mathcal{A}_+^2+\mathcal{B}_+^2)\sin^2\theta}-\frac{1}{1+\sqrt{1- (\mathcal{F} - \mathcal{C}\cos^2\frac{\theta }{2})^2 -(\mathcal{A}_+^2+\mathcal{B}_+^2)\sin^2\theta}}\right),
\eea
\end{widetext}
where $\mathcal{A},~\mathcal{B},~\mathcal{C},~ \mathcal{D},~\mathcal{F}$ are as in Eq. (\ref{eq:reduction}).
The above equation reduces to Eq.~(\ref{skew-unr}), in the absence of external noise.

\begin{figure*}[ht]
\subfigure[]{
\includegraphics[width=0.48\textwidth]{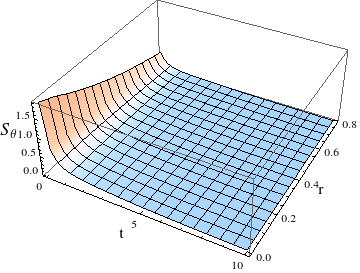}}
\subfigure[]{
\includegraphics[width=0.48\textwidth]{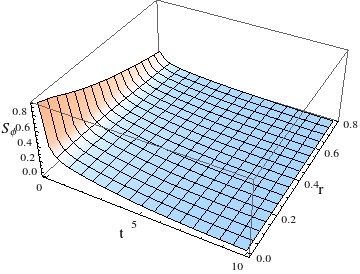}}
\caption{(a) Variation of $S_\theta$ (Skew information with respect to the  parameter $\theta$) and (b) $S_\phi$ 
(Skew information information with respect to parameter $\phi$) for SGAD interaction with bath interaction time (t) 
and and Unruh parameter (r). The parameter settings are $T=0.5$, $s=0.5$, $\theta=\pi/4$, $\phi=\pi/4$, $\phi_s=0$, $\omega_0=0.1$,             $\gamma_0=0.1$.}
\label{sgadskew-tr}
\end{figure*}

\begin{figure*}[ht]
\subfigure[]{
\includegraphics[width=0.48\textwidth]{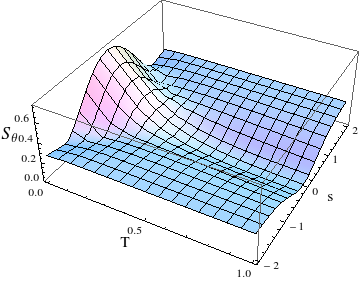}}
\subfigure[]{
\includegraphics[width=0.48\textwidth]{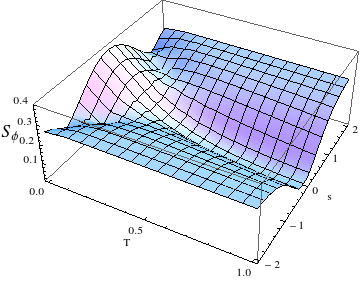}}
\caption{(a) Variation of $S_\theta$ (Skew information with respect to the  parameter $\theta$) for SGAD interaction
 with bath temperature (T) and squeezing (s). 
(b) Variation of $S_\phi$ (Skew information with respect to the  parameter $\theta$) for SGAD interaction
 with bath temperature (T) and squeezing (s).
The parameter settings are $r=\pi/8$, $\theta=\pi/4$, $\phi=\pi/4$, $\phi_s=0$, $\omega_0=0.1$, $\gamma_0=0.1$, $t=2$.}
\label{sgad-skew}
\end{figure*}

Like its Fisher counterpart, it can be seen from Fig.~\ref{sgadskew-tr}, that both $S_\theta$ and $S_\phi$ decrease with time for all values of $r$. 
The behaviour of these two Skew informations with respect to the  parameters $T$ and $s$ are depicted in Fig.~\ref{sgad-skew} (a) and (b), respectively. Qualitatively they are similar to their Fisher counterparts in Fig.~\ref{fishersgad}. Hence the rich structure exhibited by $F_\theta$ and $F_\phi$ are also seen here for their Skew counterparts.

From the behaviour of Skew information, as observed in this section, we see that it is, baring a few differences, 
quite similar to the corresponding Fisher information. This is consistent with the notion 
that the Fisher and Skew information are  variants of the same information content.

\section{Conclusions} 
Quantum Fisher information plays a prominent role in state estimation and reconstruction, 
tomography and metrology. Its variant, Skew information 
is gaining prominence in studies probing into the nature of quantum coherence. In this work, 
we provide a detailed exposition of both the Fisher and Skew information, for an Unruh-Dirac qubit. 
An important feature of this work is that by using the Bloch vector formalism, a clear and unified 
treatment of Unruh effect both in its pure form as well as in the presence of experimentally relevant 
external noise channels is provided. The use of Bloch vector representation, developed here for the Unruh effect, enables us to provide 
analytical expressions for quantum Fisher and Skew information, both with and without external noises. 
We study the evolution of Fisher and Skew information with time and also the impact of external environment parameters
such as temperature and squeezing on their evolution. The external noises are modelled by both purely dephasing phase damping as well as the squeezed generalized
amplitude damping (SGAD) noise channels. An interesting interplay between the external reservoir temperature and squeezing on the Fisher and Skew information is observed, 
in particular, for the action of the SGAD channel. It is seen that for some regimes, 
squeezing can enhance the quantum information against the debilitating influence of the noise channels. 
Similar features are also observed for the analogous study of Skew information, highlighting the similar origin of the Fisher and Skew information.
These studies, we hope, to be a contribution in the direction of efforts towards understanding and implementing 
relativistic quantum information.

\end{document}